\documentclass[twocolumn,5p]{elsarticle}
\usepackage[colorlinks=true,linkcolor=blue]{hyperref}%
\usepackage{physics}
\usepackage{graphicx}
\usepackage[T1]{fontenc}
\usepackage{ulem}
\usepackage{color}
\usepackage[utf8x]{inputenc}
\usepackage{multirow}
\usepackage{graphicx}
\usepackage{dcolumn}
\usepackage{amsmath}

\begin{document}

\title{Isotopic dependence of $(n,\alpha)$ reaction cross sections for Fe and Sn nuclei}%

\author[1]{S. K{\"u}\c{c}{\"u}ksucu}
\ead{semak@phy.hr}

\author[2]{M. Yi\u{g}it}

\author[1]{N. Paar}
\ead{npaar@phy.hr}

\affiliation[1]{organization={Department of Physics, Faculty of Science, University of Zagreb},
            addressline={Bijeni{\v c}ka c. 32}, 
            city={Zagreb},
            postcode={10000}, 
            country={Croatia}}

\affiliation[2]{organization={Department of Physics, Faculty of Arts and Science, Aksaray University},
            city={Aksaray},
            postcode={68100}, 
            country={Türkiye}}

\begin{abstract}
The $(n,\alpha)$ reactions play an important role for the energy generation and the synthesis of chemical elements in the stars, as well as for nuclear engineering and medical applications. The aim of this study is to explore the evolution of $(n,\alpha)$ reactions in Fe and Sn isotope chains in order to assess the cross section properties with
the increase of neutron number in target nucleus, and to make a comparison with other relevant neutron
induced reactions. The cross section calculations are based on the statistical Hauser-Feshbach and exciton models in TALYS nuclear reaction code, using global optical model potential that is additionally adjusted by the $(n,\alpha)$ cross section data for $^{54}$Fe and $^{118}$Sn. The calculations of  
$(n,\alpha)$ reactions in Fe and Sn isotopes provide the insight into their
isotopic dependence and properties over the complete relevant range of neutron energies. 
The cross sections result in pronounced maxima at lower-mass isotopes, and rather strong decrease for neutron-rich nuclei consistent with the reduction
of the reaction $Q$-value and increased contributions from other exit channels from compound nucleus. The analysis of the Maxwellian averaged cross sections at 
temperatures in stellar environment shows that the $(n,\alpha)$ reactions 
have significant contributions for low-mass Fe isotopes, that is opposite than for Sn isotopes. For both neutron rich isotopes $\gamma$ and neutron emissions dominate, with their interplay depending on the temperature 
involved.  
\end{abstract}

\maketitle
\section{Introduction}\label{sec:introduction}
Neutron induced nuclear reactions are essential in stellar nucleosynthesis, nuclear engineering and medical applications. The synthesis of elements heavier than Iron is governed mainly by
(i) slow neutron capture process (s-process) in giant stars \cite{Kappeler2011} and rotating massive metal poor stars \cite{Banerjee2019} and (ii) rapid neutron capture process (r-process) occurring in explosive stellar environments such as supernovae \cite{Thielemann2011} and in neutron star mergers \cite{Freiburghaus1999}. There are also other neutron capture processes like the i-process, at neutron densities between those of s- and r- process, suggested to explain part of the observed heavy element abundances \cite{Hampel2016}. 
In many astrophysical conditions $(n,\gamma)$ reactions play a dominant role, however other neutron induced reactions also participate, including the $(n,\alpha)$ reaction. In the interplay with nuclear $\beta$-decay, these reactions determine the path of the nucleosynthesis processes. Since the $(n,\alpha)$ reaction may also contribute, it is interesting to explore in more details the properties of this reaction in comparison to other competitive reactions.

The $(n,\alpha)$ reactions have recently been been investigated experimentally due to their role in nucleosynthesis, especially related to the s-process, e.g. in Refs \cite{Otuka2014,Smet2007,Gledenov2000,Goeminne2000,
Vermote2012,Weiss2014,Fotiades2015,Barbagallo2016,Gledenov2018,
Praena2018,Fe56-1,Helgesson2017,Kucuksucu2022}.
The $(n,\alpha)$ reaction has also been recently studied for $^{7}$Be target,  to address the cosmological lithium problem using the measurement of the reaction cross section in a wide energy range \cite{Barbagallo2016}.
Novel experimental techniques and detector systems have been recently developed to provide accurate new data \cite{Weiss2013,Gyurky2019}. Recent developments also include the NICE-detector,
that opened new perspectives to determine 
neutron capture cross sections with charged particle in the exit channel with 
sufficient accuracy, for different nuclear and astrophysical applications \cite{NICE2020}.
The studies of $(n,\alpha)$ reactions are also important due to their possible medical applications. For example, in the Boron Neutron Capture Therapy (BNCT) of cancer, Boron-10 is irradiated with low-energy thermal neutrons to yield high linear energy transfer particles and recoiling Lithium-7 nuclei \cite{Barth2005}. Recently, $^{33}$S has been studied as a cooperating target for the BNCT because of its large $(n,\alpha)$ cross section in the epithermal neutron energy range, that is the most suitable for the BNCT \cite{Sabate2016}.  Further studies are needed for complete description of the $(n,\alpha)$ reactions on various nuclear targets relevant in medical applications.

From the experimental side, the properties of $(n,\alpha)$ reactions remain unknown for many nuclei or they are available
in restricted neutron energy ranges \cite{JEFF3.1}. Therefore, theoretical description
is indispensable to provide complete knowledge on the cross sections relevant for many applications. 
Reaction theory of neutron capture in nuclei crucially depends on the nuclear structure and excitation properties of target and daughter nuclei involved \cite{Litvinova2009}. 
Since theoretical approaches to $(n,\alpha)$ reactions are subject to systematic 
uncertainties in various nuclear properties involved, reliable description of $(n,\alpha)$ reaction
cross sections requires both theoretical modelling and advanced experimental methods. 
The nuclear reaction cross sections, both from the experiment and model calculations are 
available in several data bases, e.g. in Refs. \cite{Otuka2014,ENDFB7.1,Rauscher2000,JEFF-3.3,JENDL4.0,ROSFOND,CENDL3.1}. Although some analyses are available on the isotopic dependence of neutron induced reactions, often they are rather limited, e.g., based on simple phenomenological formula constrained
by experimental data \cite{Zelenetskij1995}. Since the experimental data are often 
limited in terms of the incoming neutron energy, reaction theory approaches are 
essential to provide the insight how the neutron induced reaction cross sections 
evolve along the nuclide map. The advent of modern and complete nuclear reaction
theory frameworks (e.g., TALYS \cite{Koning2008,Talys}, EMPIRE\cite{Herman2007},
NON-SMOKER \cite{Rauscher2000}, etc.), opened perspective to investigate in detail 
how specific neutron induced reaction cross sections over the broad neutron energy
range evolve when going away from the valley of stability.

The aim of this work is to investigate theoretically the $(n,\alpha)$ reaction cross sections along two representative isotope chains, Fe and Sn, in order to understand how the cross sections evolve with increasing neutron number of
target nucleus.
The experimental data on $^{54}$Fe
will be used to benchmark model calculations for their extension toward
neutron rich isotopes. In order to assess the relevance of $(n,\alpha)$ reactions in 
stellar environment, neutron induced reactions with several exit channels will
be explored and compared. For this purpose, also Maxwellian averaged cross sections
 (MACS) will be calculated, assuming Maxwell-Boltzmann distribution of the incoming
 neutron flux. Model calculations will be performed using Hauser-Feshbach statistical 
 model \cite{Hauser1952,Moldauer1975} and its implementation in the nuclear reaction
 program TALYS \cite{Koning2008,Talys}. This framework represents
 one of the most advanced approaches to describe nuclear reactions, and it also
 provides the overview of the systematic uncertainties due to various inputs on
 nuclear structure and reaction properties. 

The paper is organized as follows. 
Section \ref{Sec2} includes a brief description of the theory framework and settings used 
for modelling $(n,\alpha)$ and other neutron-induced reactions of interest and its
 implementation in the TALYS code. 
In Sec. \ref{Sec31} the results on the benchmark calculation for $^{54}$Fe in comparison to the experimental data are shown. In addition, the results on the analysis of isotopic 
dependence of $(n,\alpha)$ reaction cross sections are presented in Sec. \ref{Sec31}
and Sec. \ref{Sec32} for Fe and Sn isotopes, respectively. Section \ref{Sec4}
includes the corresponding analysis of the MACS values. 
Summary and conclusions of the present work are given in Sec. \ref{Sec5}.
\section{Theory framework}\label{Sec2}

Compound nucleus reactions induced with neutrons are investigated in the
framework  of statistical Hauser-Feshbach model, while pre-equilibrium 
reaction process is described by excition model, both in their 
computational implementation in the reaction code TALYS \cite{Koning2008, Talys}. Here we give
only a brief overview of the framework settings used in this work, for more details see Ref. \cite{Talys}. Statistical model is employed to describe the binary reaction \cite{Hilaire2000}

\begin{equation}
n+A \to CN^* \to b+B,
\end{equation}

where the projectile is incoming neutron $(n)$, $A$ denotes target, $CN^*$ 
excited intermediate compound nucleus, $b$ is ejectile and B the residual nucleus.
The interaction between neutron and target nucleus $A$ is described by 
the effective (optical) potential and the reaction problem is solved numerically \cite{Hilaire2000}. While the incident channel is determined by neutrons, 
many outgoing reaction channels are possible. The compound nucleus
reaction cross sections can be described using statistical Hauser-Feshbach
model \cite{Hauser1952,Moldauer1975,Rauscher2000}. It assumes the validity
of the compound nucleus reaction mechanism and a statistical distribution of nuclear excited states \cite{Moldauer1975,Kappeler2011}. However, there are limitations
for its application, so the method is appropriate when the level density in the
contributing energy window around the peak of the projectile energy distribution is sufficiently high to justify a statistical treatment \cite{Rauscher2000}. For the energy
of incident particle below $\approx$ 20 MeV, the compound nucleus reaction
dominates. At intermediate energies, additional contributions are
possible through the pre-equilibrium reaction process, that is described by the 
exciton model \cite{Talys}. There are many possible exit channels with various outgoing particles,
where the dominant contributions come from neutron, $\gamma$-ray, proton, 
and $\alpha$-particle. The $(n,\alpha)$ reaction cross sections have been previously 
investigated with the Hauser-Feshbach model in calculations of astrophysical reaction 
rates for large set of nuclei \cite{Rauscher2000}.

For the purpose of the present work, the Hauser-Feshbach and exciton models,
implemented in the TALYS code \cite{Koning2008,Talys} are used to investigate the 
isotopic dependence of $(n,\alpha)$ reaction cross sections over the broad
energy range up to $\approx$ 20 MeV, including also a comparison with the
cross sections for several other relevant exit channels. More details about
relevant nuclear structure properties included in the model are given
in Ref. \cite{Talys}. The reaction cross sections sensitively depend
on the nuclear level densities, as well as on the optical model transmission
coefficients \cite{Talys}. 
The properties of target nuclei can be used from available experimental data or 
theoretical models.  Since the present study includes isotopic chains for which
experimental data are limited, only theoretical description can provide
all necessary quantities. Phenomenological models are often used instead of
experimental data \cite{Rauscher2000}. 
%
In this work, the Skyrme functional is used for the description of nuclear masses,
 \cite{Goriely2009}. In this way, important input for nuclear reaction study is determined from global microscopic model.

{The Hauser-Feshbach model also requires description of the transmission 
coefficients for the $\alpha$ particle emission. In this work we employ well established global optical model potential from Ref. \cite{Koning2003}, 
where the parameterisation used is given
in the TALYS implementation \cite{Talys}. However, since the
measured $(n,\alpha)$ cross sections could not be accurately described 
for two benchmark nuclei in this study, $^{54}$Fe and $^{118}$Sn, 
additional adjustments of the optical model potential has been
performed. 
More details about this procedure are given in Sec. \ref{Sec31}.
Compound nucleus cross section also includes width fluctuation correction (WFC) which accounts for the correlations between the 
incident and outgoing waves \cite{Talys}. These correlations enhance the elastic channel, and accordingly decrease other open channels. In this work the WFC factors are calculated
using standard implementation of the Moldauer model \cite{Moldauer1980}.
For the level densities, Back-shifted Fermi gas Model (BFM)
is used \cite{dilg1973level}, that fits the best the experimental data of interest
in this work.}
At intermediate incoming neutron energies the pre-equilibrium reaction channels 
start to open. Thus, for the comparison of the model calculations with 
experimental data, we also include pre-equilibrium process in theory framework. 
For the calculations of pre-equilibrium reactions, the exciton model is 
used \cite{exciton}.
With these settings, we use the TALYS code \cite{Talys} to investigate
the $(n,\alpha)$ reaction cross sections, as well as the Maxwellian averaged 
cross sections (MACS) at temperatures characteristic in stellar 
environments.

 \section{Isotopic dependence of $(n,\alpha)$ reaction cross sections}\label{Sec3}
 
By employing the nuclear reaction framework and its
computational implementation outlined in Sec. \ref{Sec2}, in the following we investigate the evolution of the properties of $(n,\alpha)$ reaction cross sections in two representative isotope chains - Fe and Sn. In addition, for comparison
we also perform more systematic analysis including other relevant 
exit channels, including neutron, proton, and $\gamma$ emission.   
 
\subsection{Fe isotopes}\label{Sec31}

First we consider model calculations for ${}^{54}$Fe, because it is  the only Fe isotope with available systematic experimental data on $(n,\alpha)$  reaction cross sections covering a broad range of the incoming neutron  energies up to $\approx$ 18 MeV \cite{expFe54-1,expFe54-2,expFe54-3,expFe54-4,expFe54-5,
expFe54-6,expFe54-7,expFe54-8,expFe54-9,expFe54-10,
expFe54-11,expFe54-12,expFe54-13}.
Therefore, we choose ${}^{54}$Fe for benchmarking the settings of the nuclear
reaction model for further studies in Fe isotopes. Since we aim to study 
the systematics of the cross sections along isotope chains for $(n,\alpha)$ reactions, the most appropriate are the implementation of the global optical model 
potential of Koning and Delaroche \cite{Koning2003} and alpha optical model from Avrigeanu et al. \cite{Avrigeanu2014} as outlined in Ref. \cite{Talys}.
The $\alpha$-nucleus optical potential is usually constrained by employing data from 
elastic alpha scattering experiments \cite{PhysRev.107.1343,PhysRev.106.126,PhysRev.95.1212,AVRIGEANU2003104,MOHR2013651,PhysRevC.85.035808}.
In order to improve description of experimental data for $(n,\alpha)$ cross sections for ${}^{54}$Fe \cite{expFe54-1,expFe54-2,expFe54-3,expFe54-4,expFe54-5,expFe54-6,expFe54-7,expFe54-8,expFe54-9,expFe54-10,expFe54-11,expFe54-12,expFe54-13}, a rescaling of the optical model potential mass dependent geometry parameters for radius $r_{v}$, $r_{w}$, for 
diffuseness $a_{v}$, $a_{w}$ in the volume central potential and corresponding 
parameters $r_{D}$, $a_{D}$ in the surface central potential are considered \cite{Koning2003}.
To optimize global parameterization of alpha optical model potential \cite{Avrigeanu2014}, 
we performed $\chi^{2}$ minimization of the cross sections for $(n,\alpha)$ reaction 
on ${ }^{54}$Fe using MINUIT package \cite{minuit} and available experimental data
mentioned above.
In this way, we obtained the rescaling
factors for the geometrical parameters of the global optical model potential from Ref. \cite{Avrigeanu2014}, 
as given in Table \ref{tableparameters}.
\begin{table}[h!]
\caption{Rescaling factors for the geometrical parameters in global alpha optical 
model potential \cite{Avrigeanu2014} for ${}^{54}$Fe and ${}^{118}$Sn.}
\centering
\begin{tabular}{ c c c } 
\hline
       &  ${}^{54}$Fe & ${}^{118}$Sn  \\
\hline
 $r_{v}$& 0.94658 & 0.95283 \\
 $a_{v}$ & 1.05843 & 0.71722  \\ 
 $r_{w}$ & 0.80276 & 1.20286   \\ 
 $a_{w}$ & 1.19983 & 0.70011  \\ 
 $r_{D}$ & 1.01336 & 1.04916 \\ 
 $a_{D}$ & 1.09999 & 1.05427 \\ 
 \hline
\end{tabular}
\label{tableparameters}
\end{table}

The $\chi^{2}$ values using adjusted optical model potential parameters (TALYS-1.96(a)) are shown in Table \ref{tablechi} in comparison to those of default TALYS-1.96 calculation and
the corresponding $\chi^{2}$ values obtained for a selection of other theoretical approaches, tabulated in JEFF-3.3 \cite{JEFF-3.3}, ENDF/B-VII.1 \cite{ENDFB7.1}, 
TENDL-2019 \cite{TENDL-2019}, BROND-3.1 \cite{BROND3.1}, CENDL-3.2 \cite{CENDL3.2}, JENDL-5 \cite{JENDL-5}, ROSFOND-2010 \cite{ROSFOND} and NON-SMOKER \cite{Rauscher2000} data sets. The rescaling of the global optical model parameters provides improvement for the description of the $(n,\alpha)$ cross sections and in this way we benchmark the theory framework for further study of the Fe isotope chain.
\begin{table}[h!]
\caption{The $\chi^{2}$ values for $(n,\alpha)$ cross sections for ${}^{54}$Fe and ${}^{118}$Sn in the present TALYS-1.96 calculation, in comparison to those calculated with the cross sections from other calculations. TALYS-1.96(a) denotes optimized calculation after adjustment of optical model parameters (see text for details).}
\centering
\begin{tabular}{ c c c } 
\hline
       &  $\chi^2$ (${}^{54}$Fe) & $\chi^2$ (${}^{118}$Sn)  \\
\hline
 TALYS-1.96 & 3.14 & 27.606 \\
 TALYS-1.96(a) & 2.61 & 11.65  \\ 
 JEFF-3.3 \cite{JEFF-3.3} & 7.03 & 79.84   \\ 
 ENDF/B-VII.1 \cite{ENDFB7.1} & 2.74 & 12.47  \\ 
 TENDL-2019 \cite{TENDL-2019} & 1.46 & 156.18 \\ 
 BROND-3.1 \cite{BROND3.1} & 3.18 & 12.47  \\ 
 CENDL-3.2 \cite{CENDL3.2} & 2.86 & 75.14 \\
 JENDL-5 \cite{JENDL-5} & 2.06 & 81.10 \\
 ROSFOND-2010 \cite{ROSFOND} & 7.09 & 12.41 \\
 NON-SMOKER \cite{Rauscher2000} & 19.83 & 6426.48 \\
 \hline
\end{tabular}
\label{tablechi}
\end{table}

\begin{figure}
\centering
\includegraphics[width=\linewidth]{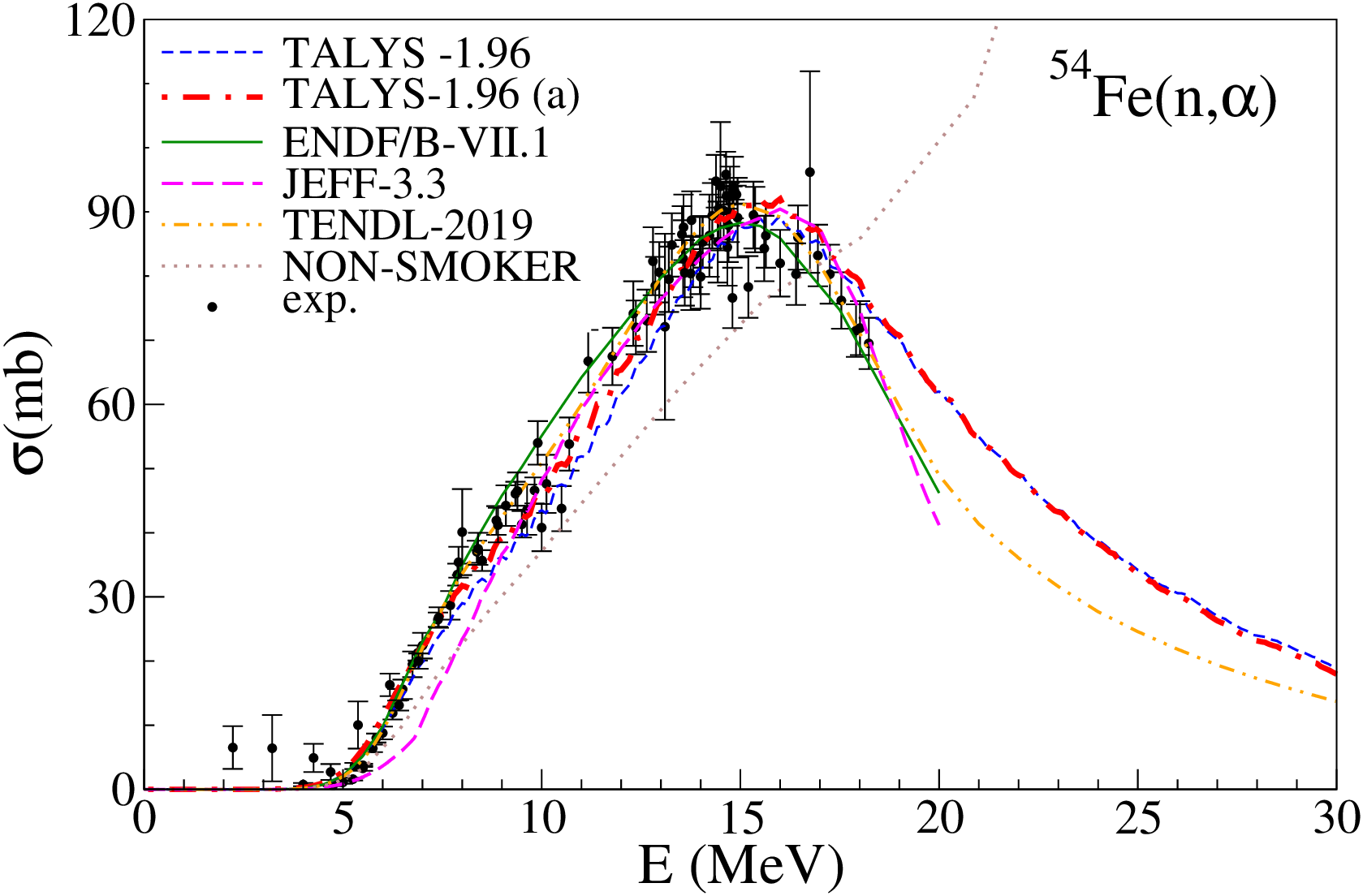}
\caption{The $(n,\alpha)$ reaction cross section for ${}^{54}$Fe from the present TALYS calculation in comparison with the experimental data \cite{expFe54-1,expFe54-2,expFe54-3,expFe54-4,expFe54-5,
expFe54-6,expFe54-7,expFe54-8,expFe54-9,expFe54-10,
expFe54-11,expFe54-12,expFe54-13} and various model calculations for ENDF/B-VII.1 \cite{ENDFB7.1}, JEFF-3.3 \cite{JEFF-3.3}, TENDL-2019 \cite{TENDL-2019} and reaction code NON-SMOKER \cite{Thielemann2011}. TALYS-1.96(a) denotes optimized calculation after adjustment of optical model parameters.}\label{Fe54libs1}
\end{figure}

\begin{figure}
\centering
\includegraphics[width=\linewidth]{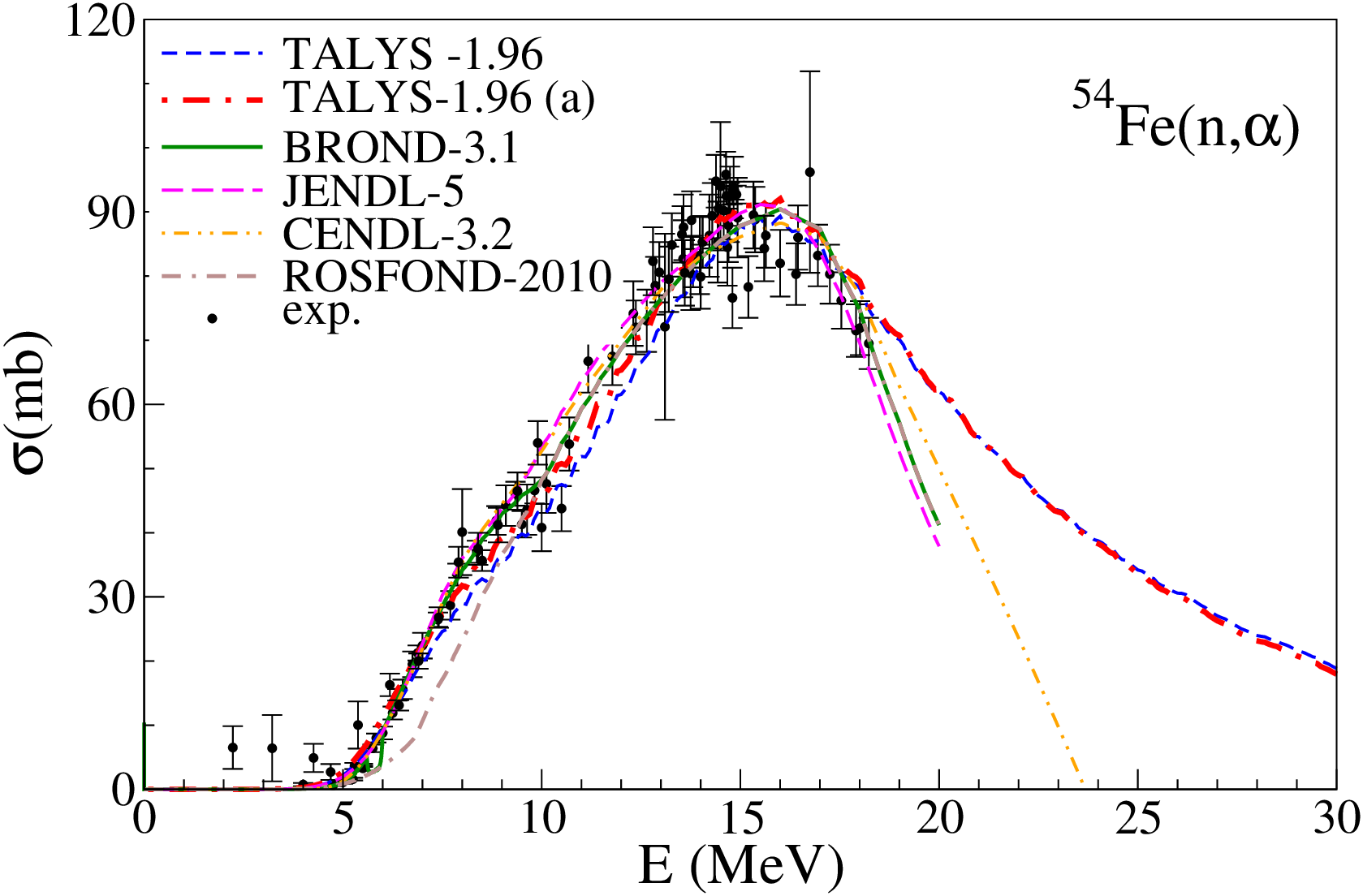}
\caption{The same as Fig. \ref{Fe54libs1} but BROND3.1 \cite{BROND3.1}, JENDL-5 \cite{CENDL3.2} and ROSFOND \cite{Rauscher2010} model calculations are shown for comparison.}\label{Fe54libs2}
\end{figure}

The results of the present calculation of $(n,\alpha)$ reaction cross sections for ${}^{54}$Fe are shown in Fig.~\ref{Fe54libs1} and Fig.~\ref{Fe54libs2} over the complete energy range up to 30 MeV.
The cross sections are compared with
the collection of experimental data from various studies \cite{expFe54-1,expFe54-2,expFe54-3,expFe54-4,expFe54-5,
expFe54-6,expFe54-7,expFe54-8,expFe54-9,expFe54-10,
expFe54-11,expFe54-12,expFe54-13},
as well as those of other calculations. We have restricted the 
comparison to the results summarized in Refs. ENDF/B-VII.1
\cite{ENDFB7.1}, JEFF-3.3 \cite{JEFF-3.3}, TENDL-2019 \cite{TENDL-2019}, NON-SMOKER \cite{Rauscher2000}, BROND-3.1 \cite{BROND3.1},  JENDL-5 \cite{JENDL-5}, CENDL-3.2 \cite{CENDL3.2}, and ROSFOND-2010 \cite{ROSFOND}. The results of the present model calculation, displaying the
Gaussian-like shape of the cross sections characteristic for the 
compound nucleus reactions with particle emission, are in excellent agreement with the experimental data, that is improved in comparison to the 
previous studies.
The maximal value of the cross section is obtained at neutron energy E= 15.62 MeV, that is very close to the peak energy from the experimental data as shown in Fig.~\ref{Fe54libs1} and Fig.~\ref{Fe54libs2}.
Although we do not consider low-energy cross sections in detail, we note that
the largest sensitivity of the statistical model calculation of 
$(n,\alpha)$ cross sections corresponds to the lowest incident energies,
where the level-density effects are not present \cite{Kucuksucu2022}. For example, the cross sections 
in the present TALYS-1.96 and TALYS-1.96(a) calculations at energies below 4 MeV differ up to several orders of magnitude. 

\begin{figure}
\centering
\includegraphics[width=\linewidth]{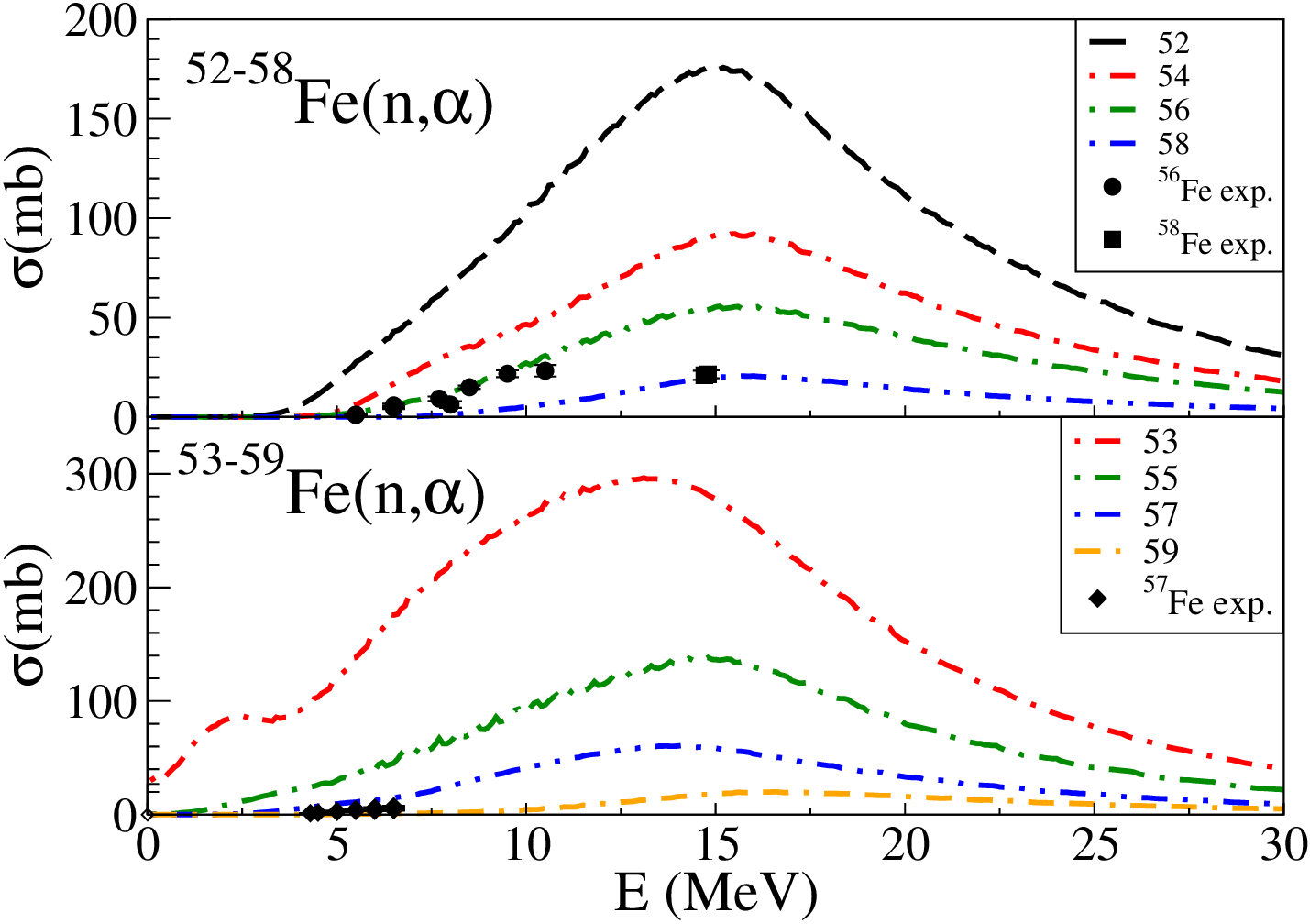}
\caption{The $(n,\alpha)$ reaction cross sections for even-even isotopes $^{48-58}$Fe (top panel) and even-odd isotopes $^{49-59}$Fe (lower panel), based on TALYS-a calculation. The experimental cross sections are also shown
for $^{56}$Fe \cite{Fe56-1,expFe54-5,expFe54-8,Grimes1979,Kneff1981,Kneff1996,Fischer1984,Sterbenz}, $^{57}$Fe \cite{expFe54-10,Fe57-2}, and $^{58}$Fe \cite{Fe58-1,Fe58-2}.}\label{Fe-isotopes}
\end{figure}

\begin{figure}
\centering
\includegraphics[width=\linewidth]{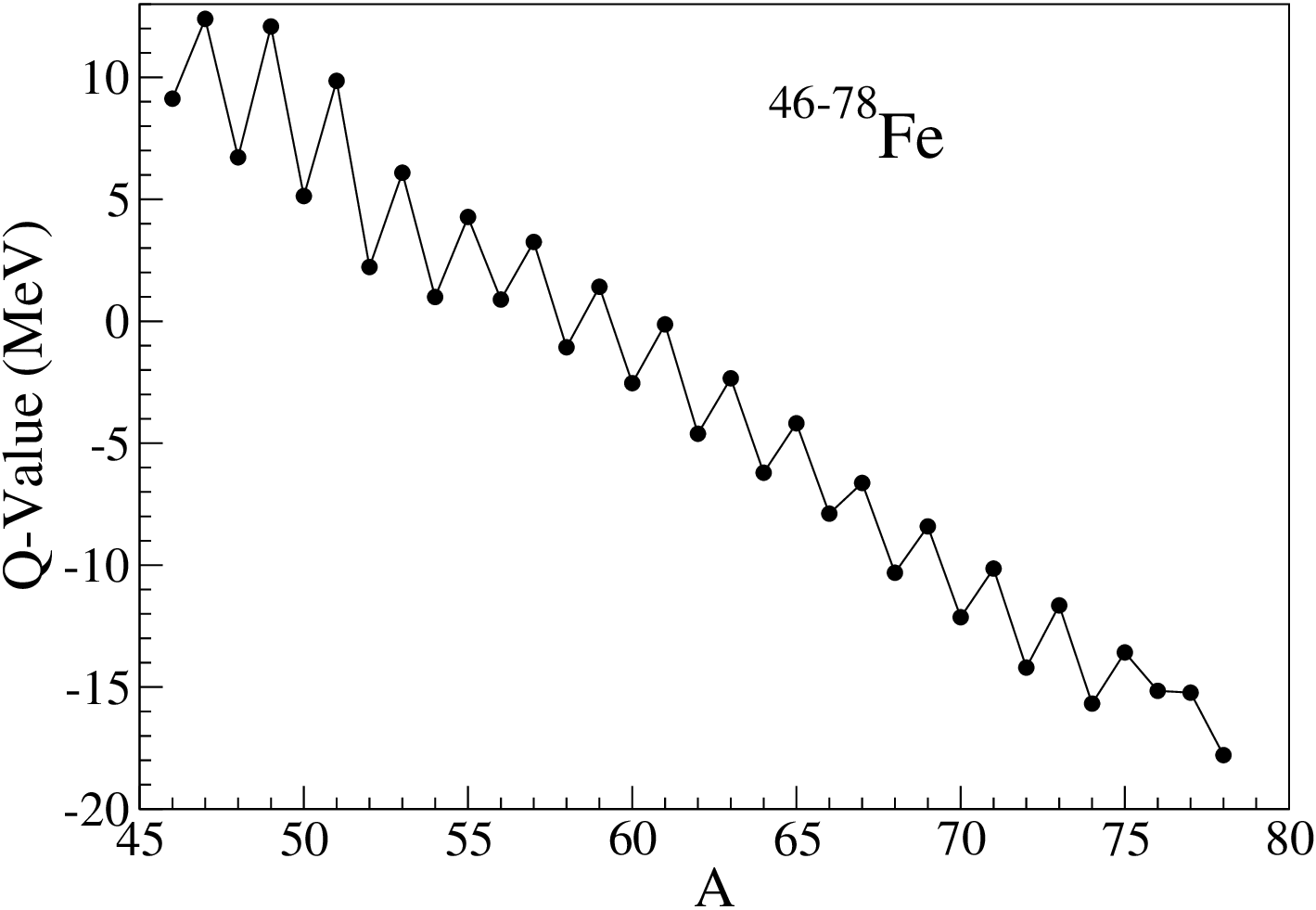}
\caption{The $(n,\alpha)$ reaction $Q$-value for Fe isotopes.}
\label{QvalueFe}
\end{figure}

We employ the theory framework benchmarked on the experimental data 
for ${}^{54}$Fe in the study of the evolution of the $(n,\alpha)$ reaction cross sections for Fe isotope chain. The results for 
the cross sections are shown in Fig. \ref{Fe-isotopes} for 
even-even isotopes $^{48-58}$Fe (top panel) and even-odd isotopes $^{49-59}$Fe (lower panel). For comparison, the experimental cross sections are also shown
for a few additional isotopes with experimental data available, $^{56}$Fe \cite{Fe56-1,expFe54-5,expFe54-8,Grimes1979,Kneff1981,Kneff1996,Fischer1984,Sterbenz}, $^{57}$Fe \cite{expFe54-10,Fe57-2}, and $^{58}$Fe \cite{Fe58-1,Fe58-2}. Excellent agreement of the 
calculated cross sections with experimental data for $^{56,57,58}$Fe is obtained, 
without any additional adjustments.
Thus, the established model settings seem to be appropriate to study the systematics of the cross
sections along the Fe isotope chain. 
For even-even Fe isotopes, with further increase of the 
neutron number, the cross sections become smaller. Similar behaviour 
can be observed for even-odd Fe isotopes. In order to understand this dependence of the
cross sections, it is interesting to inspect the corresponding reaction 
$Q$-values along the isotope chain, as shown in Fig. \ref{QvalueFe}.
The $Q$-values systematically decrease with the number of neutrons, 
showing also odd-even staggering, thus, for more neutron rich isotopes
more energy is required for the reaction. Emission of the $\alpha$ particle
from the compound nucleus is determined by the probability for its
formation within nucleus, and transmission probability for tunneling
through the Coulomb barrier. The reaction cross section depends 
on the compound nucleus Coulomb barrier, since it suppresses the emission 
of outgoing $^{4}$He. Although the height of the Coulomb barrier does not 
change along the isotope chain, it becomes spatially extended toward 
larger values of radial coordinate \cite{Lisboa2004}. Finally, with 
increasing the neutron number, other particle emission channels also become 
more active, thus the $\alpha$ emission
becomes reduced. We will discuss this aspect in more details.

\begin{figure}
\centering
\includegraphics[width=\linewidth]{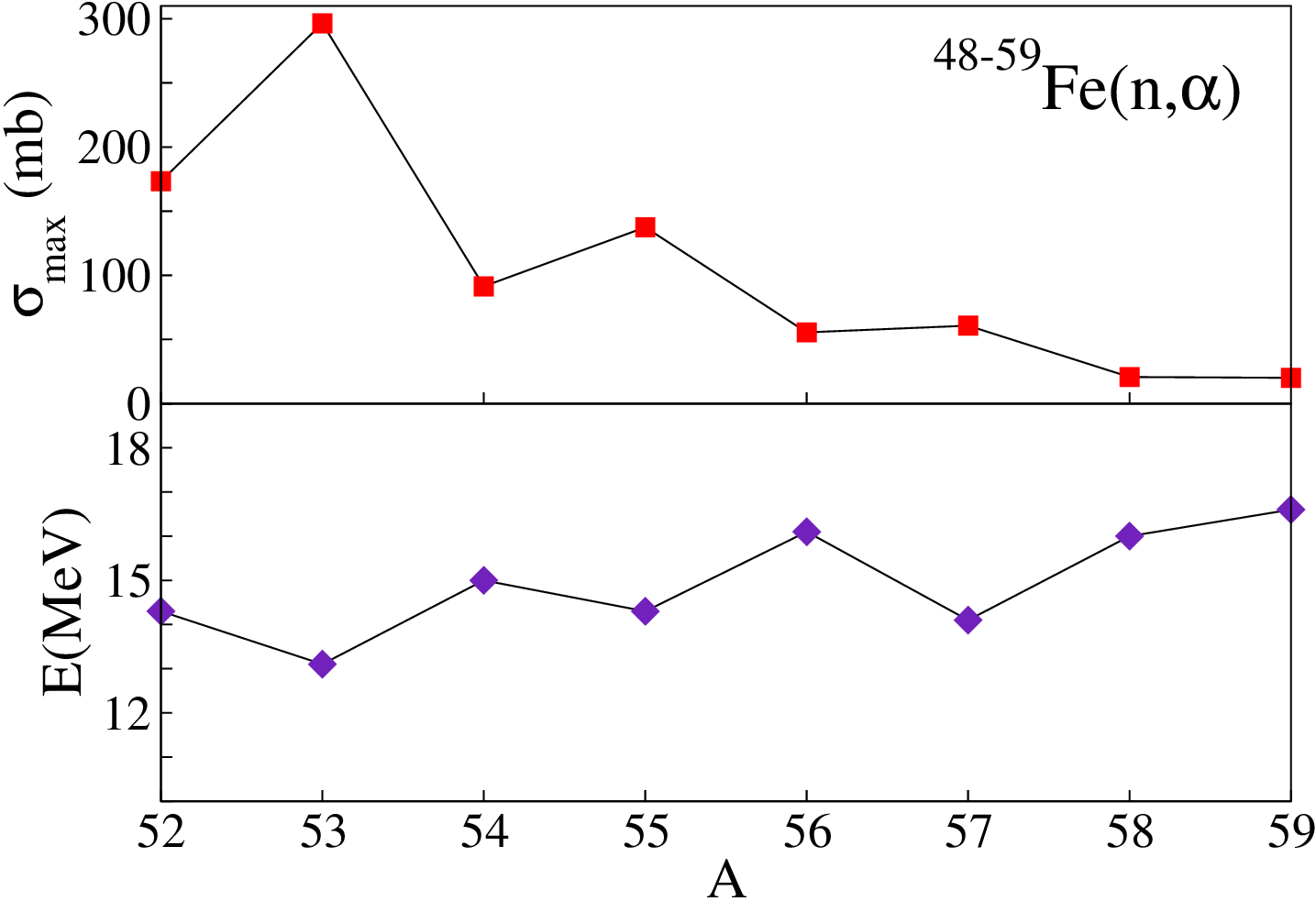}
\caption{The maximal cross section value for $(n,\alpha)$ reaction for ${}^{46-67}$Fe (upper panel) and the corresponding energy of the maximal cross section (lower panel).}\label{Fe-sigma-max}
\end{figure}

Next we analyze the properties of the main peak of the cross section.
Figure \ref{Fe-sigma-max} shows the maximal value of the cross section
and the corresponding neutron energy. 
The maximum of the
cross section along the Fe isotope chain displays behaviour previously discussed,
after reaching the largest value for $^{53}$Fe, it is
rapidly decreasing by almost two orders of magnitude for $^{59}$Fe. The energy of 
the reaction cross section peak for $N>Z$ isotopes 
converges around $\approx$ 15 MeV. This saturation of the neutron energy required
for the maximal $(n,\alpha)$ reaction cross section in
isotopes with neutron excess appears as interesting feature of this reaction.
The decrease of the cross section from its maximal value is in
agreement with the isotope effect originally postulated in Ref.\cite{Gardner1964,Gardner1967}
and confirmed in Ref.\cite{Molla1977} on the basis of $Q$-values and 
the stable-isotope data. 
Rather constant energy around 15
MeV of the excitation-function maxima for stable isotopes also explains
the former establishment of the isotope effect at this energy.
 
\begin{figure}
\centering
\includegraphics[width=\linewidth]{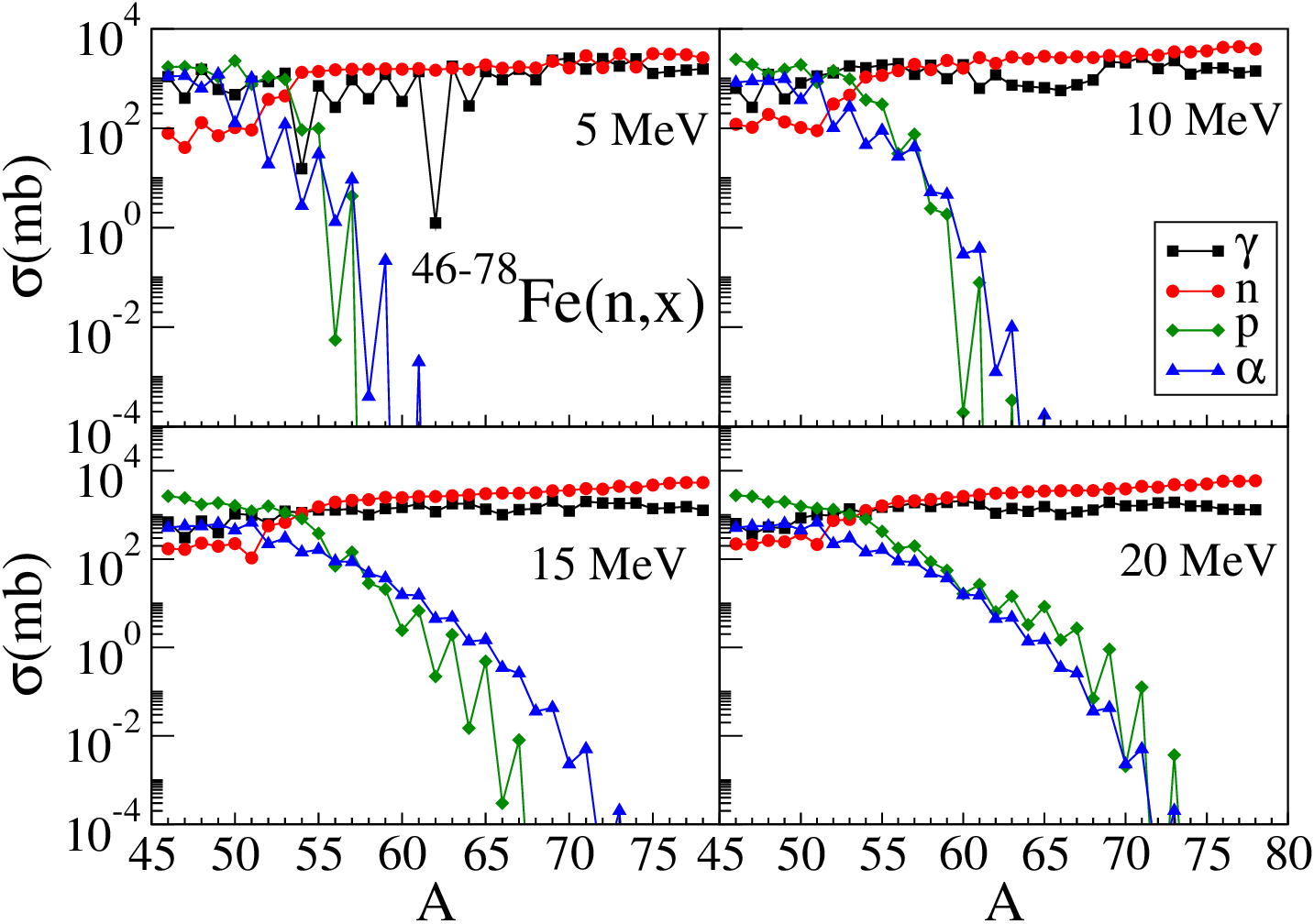}
\caption{The cross sections for neutron induced reactions with ${}^{46-78}$Fe target for gamma, neutron, proton and alpha emission from the compound nucleus, 
shown at 5, 10, 15 and 20 MeV neutron energies.\label{Fe-5-10-15-20}}
\end{figure}

As mentioned before, one of the reasons for reducing the $(n,\alpha)$ reaction cross
sections along the isotope chain are increasing contributions from other 
exit channels. For this purpose, in Fig. \ref{Fe-5-10-15-20} the cross sections are shown for the neutron induced reactions with Fe isotopes for 5, 10, 15 and 20 MeV neutron energies and various exit channels, including $\gamma$, $n$, $p$, and $\alpha$ emission from the compound nucleus. For ${}^{46-53}$Fe at all neutron 
energies the cross sections moderately increase, except for protons, 
showing odd-even staggering due to sensitivity on the pairing correlations. In the same mass region the cross
section for the proton emission decreases, while the one for neutron emission
increases, as expected for the increase of the neutron number in target nucleus.
Of course, compound nucleus with excess of neutrons will favor emission of
neutrons rather than protons, besides, neutrons are not affected by the
Coulomb barrier.
One can also observe an overall increase of the cross section for the gamma emission
in the exit channel. Beyond ${}^{53}$Fe the cross sections for $\alpha$, as well as proton emission from the compound nucleus rapidly decrease, i.e. the corresponding
cross sections reduce by several order of magnitude when compared to other shown
exit channels. Clearly, for neutron rich Fe isotopes the
$(n,\alpha)$ reaction cross section becomes negligible.

\subsection{Sn isotopes}\label{Sec32}

We extend our analysis of the $(n,\alpha)$ reaction cross sections to
medium heavy mass region, for Sn isotopes as target nuclei. The
cross section experimental data are rather limited and available over 
broader, though restricted, range of energies only for ${}^{118}$Sn \cite{Sn118exp1,Sn118exp2,Sn118exp3,Sn118exp4,Sn118exp5,Sn118exp6,
Sn118exp7,Sn118exp8}. 
Similar as for Fe isotopes, first we adjust the settings of the model
by using experimental data for ${}^{118}$Sn. 
We constrain the parameters of the alpha global 
optical model potential, $r_{v}$, $a_{v}$, $r_{w}$, $a_{w}$, $r_{D}$, $a_{D}$, by
$\chi^{2}$ minimization of the $(n,\alpha)$ cross sections
for ${}^{118}$Sn using the experimental data from Refs. \cite{Sn118exp1,Sn118exp2,Sn118exp3,Sn118exp4,Sn118exp5,Sn118exp6,
Sn118exp7,Sn118exp8}.
The resulting rescaling factors of the global optical model potential 
from Ref. \cite{Avrigeanu2014} are shown in Table \ref{tableparameters}.
Furthermore, Table \ref{tablechi} shows the $\chi^{2}$ value obtained by adjusting the
optical model parameters in the present TALYS calculation, in comparison
to the corresponding $\chi^{2}$ values obtained from other calculations 
\cite{JEFF-3.3,Rauscher2000,ENDFB7.1,TENDL-2019,BROND3.1,CENDL3.2,JENDL-5,ROSFOND}
The present adjustment
of the optical model results in improved description of the $(n,\alpha)$ cross sections, and in this way we benchmark the model for systematic studies of the cross sections along Sn isotope chain.

\begin{figure}
\centering
\includegraphics[width=\linewidth]{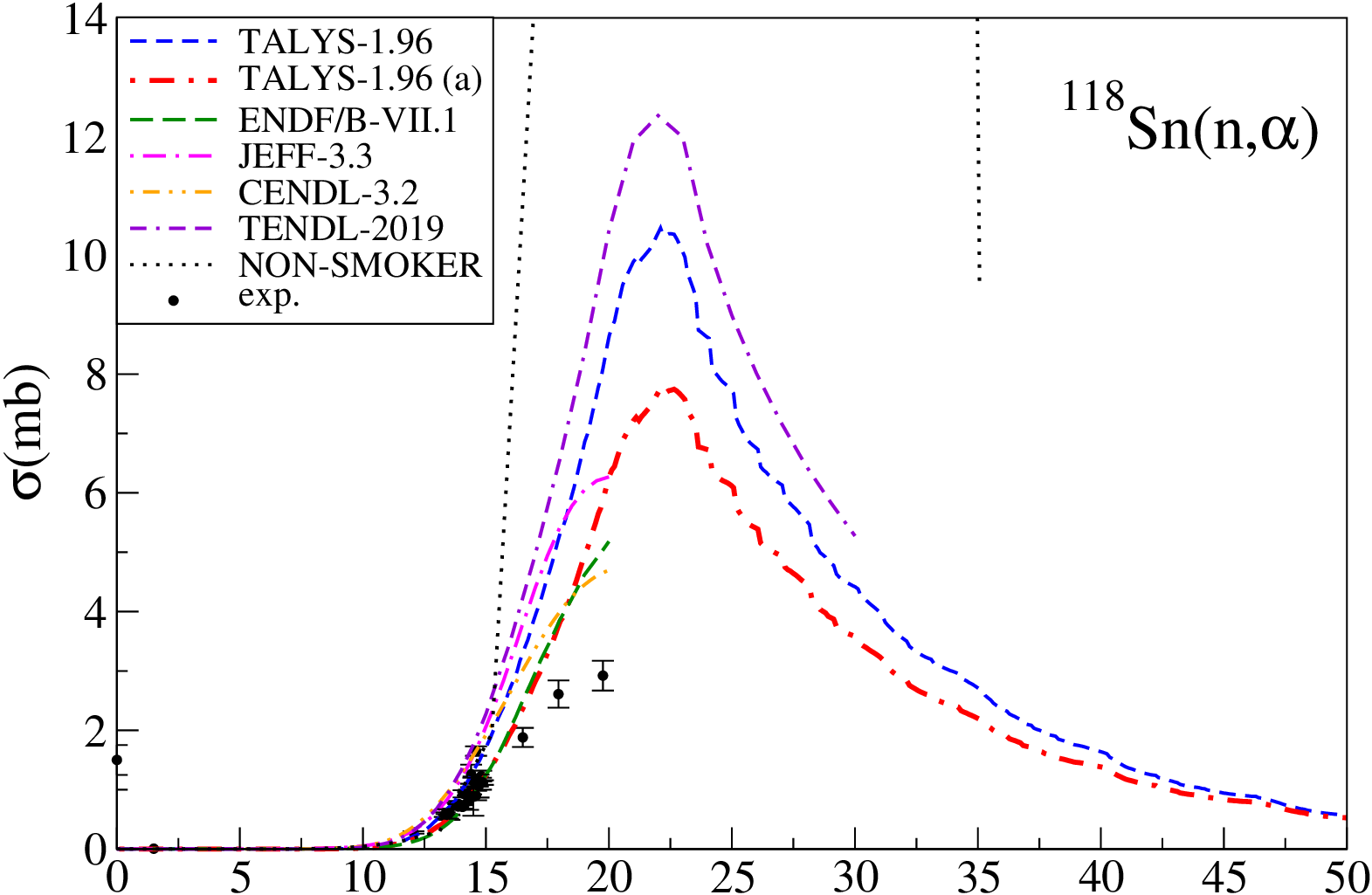}
\caption{The $(n,\alpha)$ reaction cross sections for ${}^{118}$Sn. The present TALYS-a calculation is compared with experimental data \cite{Sn118exp1,Sn118exp2,Sn118exp3,Sn118exp4,Sn118exp5,Sn118exp6,
Sn118exp7,Sn118exp8} and other theoretical approaches  \cite{JEFF-3.3,ENDFB7.1, CENDL3.2,TENDL-2019,Rauscher2000}.}\label{Sn118libs1}
\end{figure}

\begin{figure}
\centering
\includegraphics[width=\linewidth]{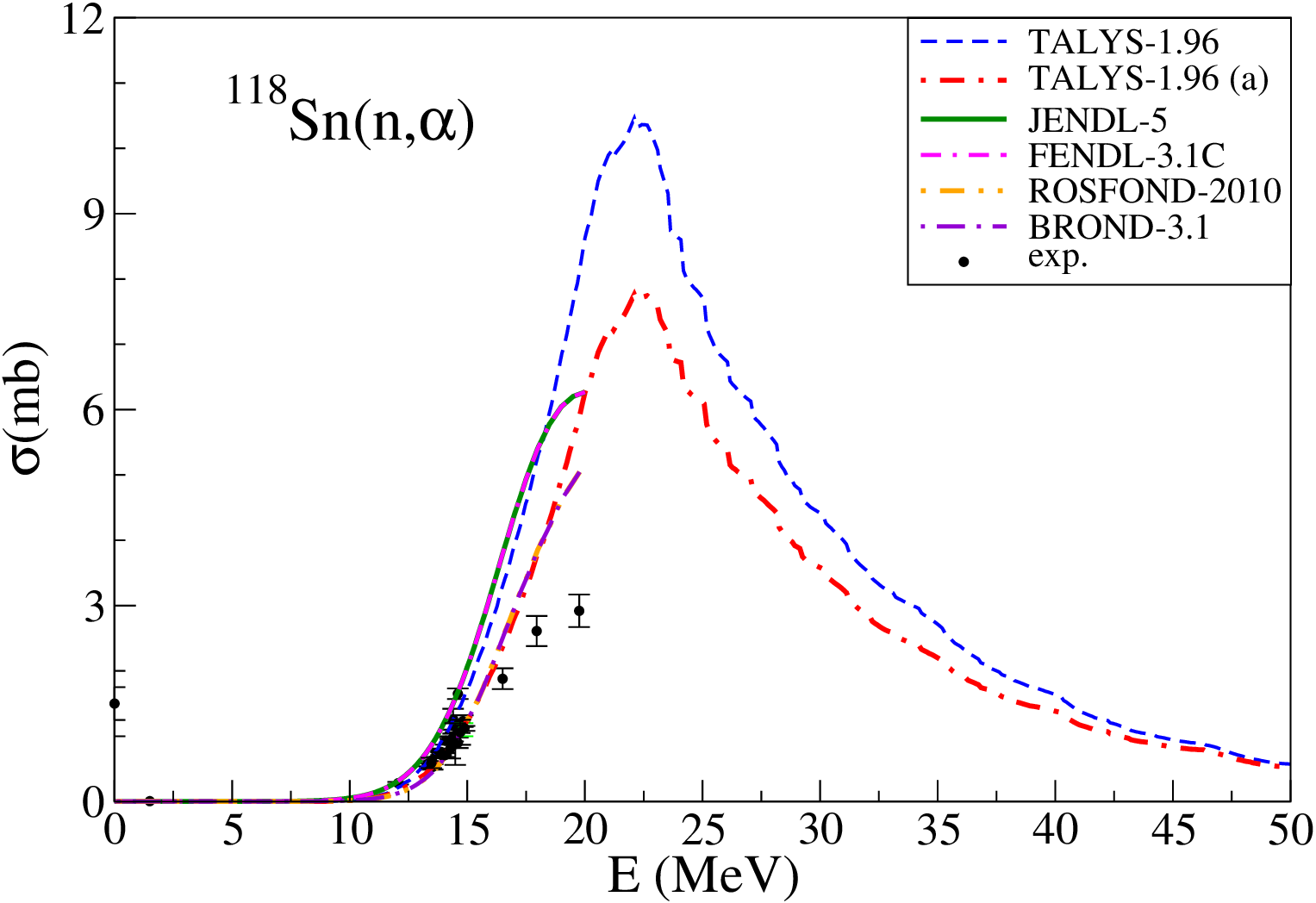}
\caption{The $(n,\alpha)$ reaction cross sections for ${}^{118}$Sn. The present TALYS-a calculation is compared with experimental data \cite{Sn118exp1,Sn118exp2,Sn118exp3,Sn118exp4,Sn118exp5,Sn118exp6,
Sn118exp7,Sn118exp8} and other theoretical approaches. \cite{JENDL-5,BROND3.1,ROSFOND,JENDL-5,FENDL-3}.}\label{Sn118libs2}
\end{figure}

The resulting $(n,\alpha)$ cross sections for ${}^{118}$Sn are shown in Fig. \ref{Sn118libs1} and Fig. \ref{Sn118libs2} , in comparison to experimental data \cite{Sn118exp1,Sn118exp2,Sn118exp3,Sn118exp4,Sn118exp5,Sn118exp6,
Sn118exp7,Sn118exp8} and calculations from Refs. \cite{JEFF-3.3, ENDFB7.1, JENDL-5, CENDL3.2, TENDL-2019, BROND3.1, ROSFOND, FENDL-3, JENDL-5, Rauscher2000}.
Due to limited data, from the experimental side it is not possible to observe
the complete shape of the cross section, while the present TALYS calculation 
results in a Gaussian-like shape of the cross section peaked around 22.27 MeV.
In comparison to other studies, that often have restricted energy range for
the cross sections, the present calculation represents considerable 
improvement, especially when compared to NON-SMOKER \cite{Rauscher2000} and
TENDL-2019 \cite{TENDL-2019}. However,
for a complete understanding of the $(n,\alpha)$ cross sections, more
experimental data are required, covering a broader range of neutron 
energies, as shown in Fig. \ref{Sn118libs1} and Fig. \ref{Sn118libs2}.

\begin{figure}
\centering
\includegraphics[width=\linewidth]{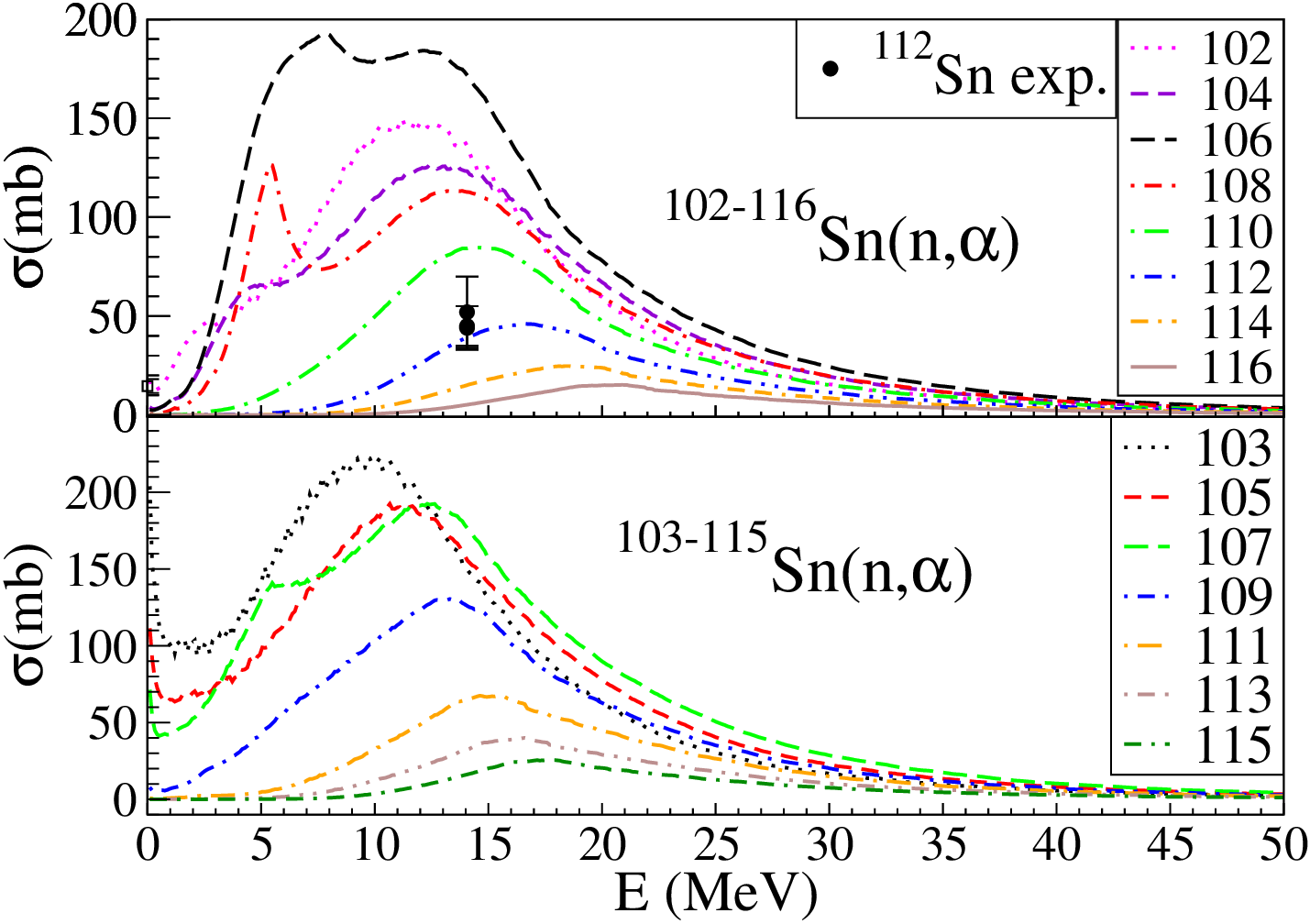}
\caption{The $(n,\alpha)$ reaction cross sections for ${}^{102-116}$Sn, shown separately for even-even (upper panel) and even-odd (lower panel) 
isotopes. The experimental cross section for $^{112}$Sn is also 
shown \cite{Sn112-1}.}\label{Sn-isotopes}
\end{figure}

\begin{figure}
\centering
\includegraphics[width=\linewidth]{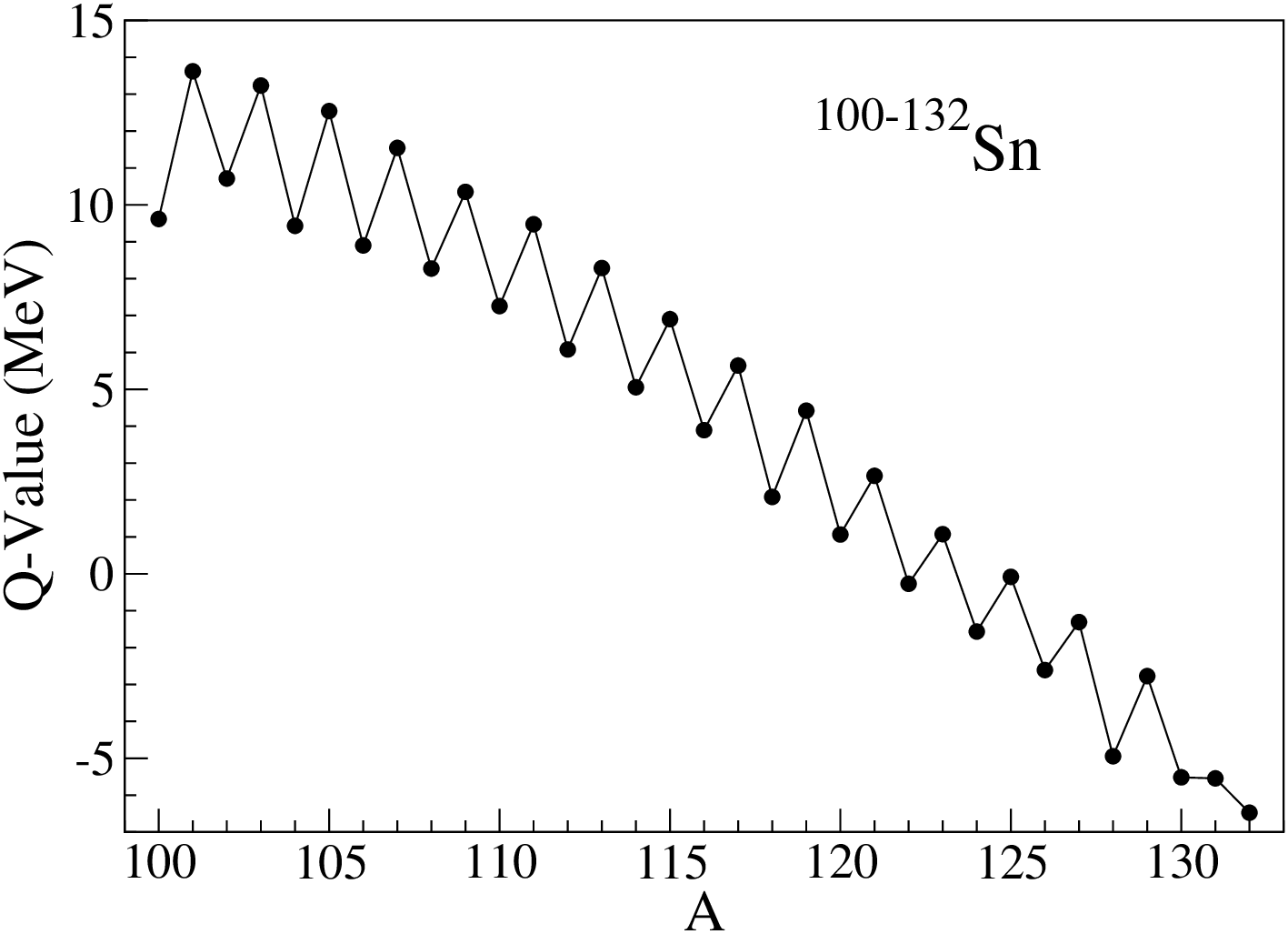}
\caption{The $(n,\alpha)$ reaction $Q$-value for Sn isotopes.}
\label{QvalueSn}
\end{figure}

By using TALYS calculation with previously established rescaling for the
optical model parameters, in Fig. \ref{Sn-isotopes} the 
$(n,\alpha)$ cross sections are shown for the range of isotopes  ${}^{102-116}$Sn, 
separately for even-even and even-odd nuclei. In the former case,
the overall cross sections reach their maximal values for ${}^{106}$Sn, and they 
further systematically decrease till ${}^{116}$Sn, resulting
with very small values, e.g. its maximal cross section at $E=$20.68 MeV is 
only 16.01 mb, in comparison to ${}^{106}$Sn at $E=$7.8 MeV, where $\sigma=$194.17mb. 
As shown in the lower panel of Fig. \ref{Sn-isotopes}, similar behaviour of the
cross sections is also obtained for even-odd Sn isotopes, with the maximal
cross section obtained for ${}^{103}$Sn. Qualitatively, with increasing
the neutron number, we obtain similar behaviour of the $(n,\alpha)$ 
cross sections as previously shown for Fe isotopes, and the same 
discussion on their isotopic dependence also applies for Sn isotopes. 
The dependence of the cross sections is also determined by the
reaction $Q$-value, which systematically decreases along the Sn isotope
chain, with odd-even staggering, as shown in Fig. \ref{QvalueSn}.

\begin{figure}
\centering
\includegraphics[width=\linewidth]{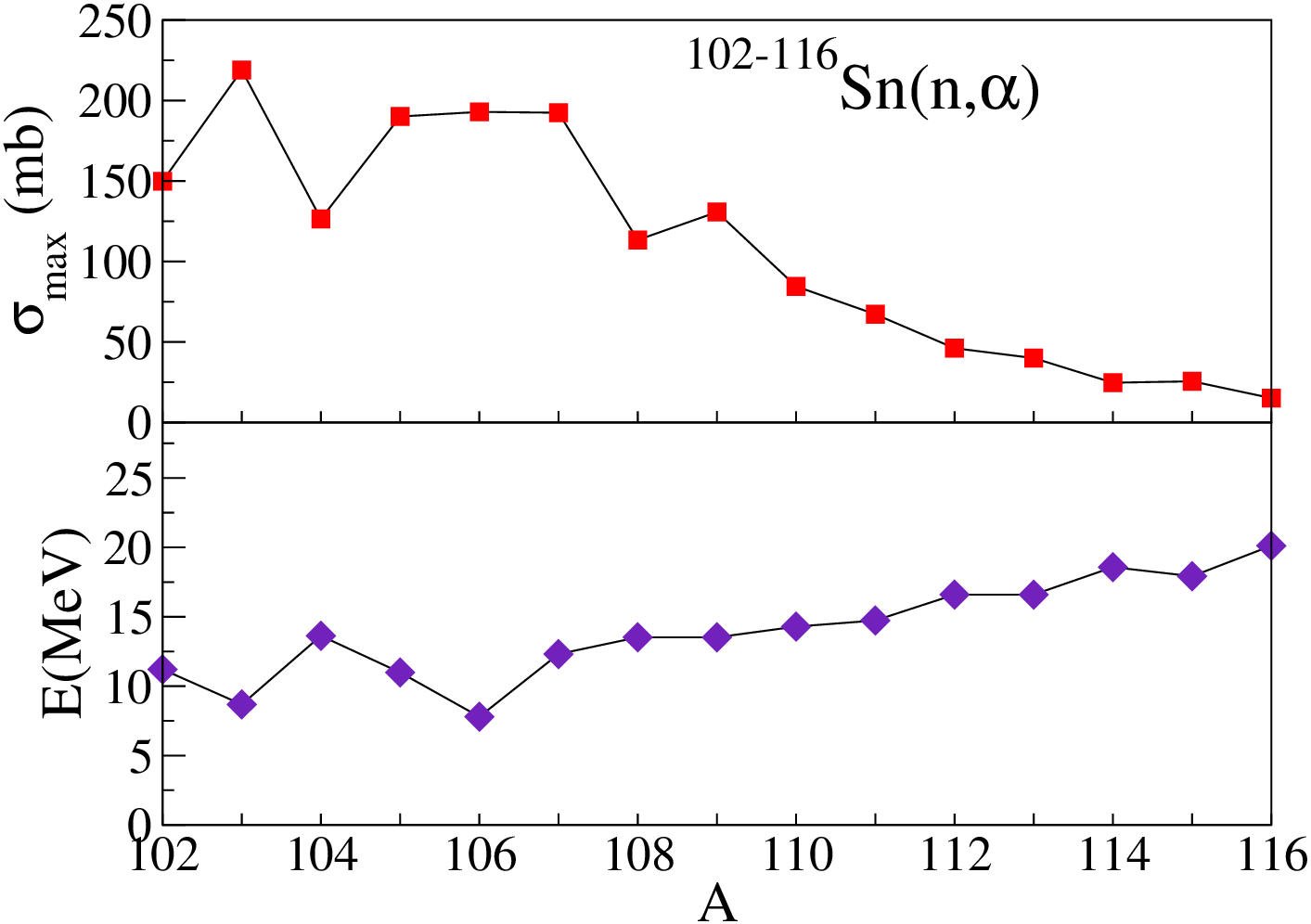}
\caption{The same as Fig. \ref{Fe-sigma-max} but for ${}^{102-116}$Sn isotopes.}
\label{Sn-sigma-max}
\end{figure}

To analyse the systematic behaviour of the $(n,\alpha)$ cross sections for Sn isotopes,
in Fig. \ref{Sn-sigma-max} the maximal values or the cross sections and the corresponding 
neutron energies are shown for ${}^{102-116}$Sn. The maximal cross section peaks are obtained
for $^{106}$Sn (even-even isotopes) and $^{103}$Sn (even-odd isotopes). 
With further increase of the neutron number, more energy
of the incoming neutrons is required for the reaction, in agreement with the evolution of
the $Q$-value with the neutron number, as shown in Fig. \ref{QvalueSn}. Lower panel of
Fig. \ref{Sn-sigma-max} also shows the saturation of the neutron energy of the maximal 
cross section with increase of the neutron number, though slight increase of the cross section
with the energy is obtained.

\begin{figure}
\centering
\includegraphics[width=\linewidth]{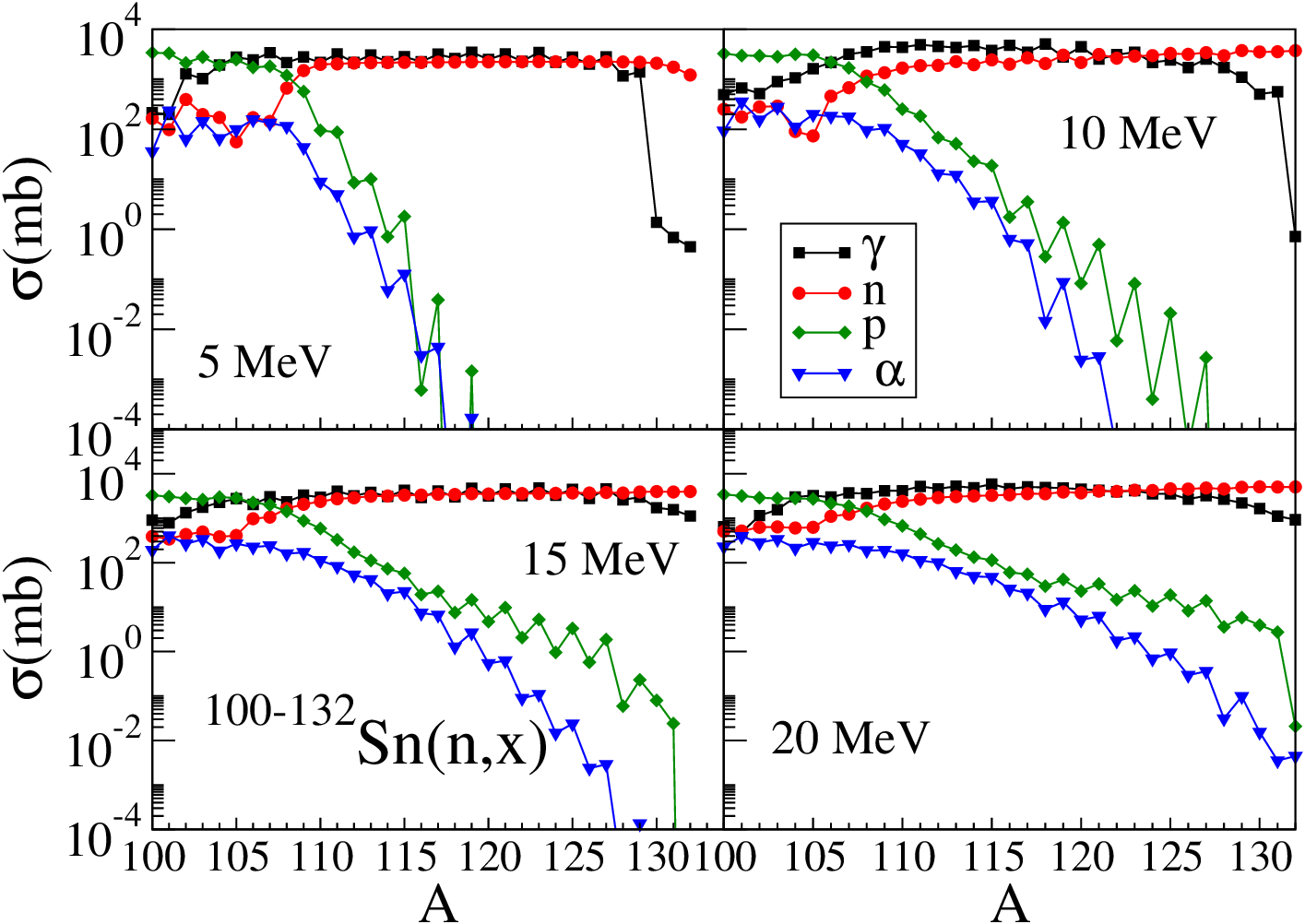}
\caption{The same as Fig. \ref{Fe-5-10-15-20}, but for ${}^{100-132}$Sn isotopes.}
\label{Sn-5-10-15-20}
\end{figure}

Figure \ref{Sn-5-10-15-20} shows the evolution of the cross sections for neutron induced
reactions for $^{100-132}$Sn at neutron energies 5,10,15, and 20 MeV, by considering
$\gamma$, $n$, $p$, and $\alpha$ exit channels. In general, the cross sections for $p$ and $\alpha$ emission
systematically reduce with the neutron number, while those for the neutron and gamma emission increase or saturate, as expected for compound nuclei with larger number of neutrons. Some deviations are obtained for gamma emission involving nuclei with large neutron excess. The reduction of
$(n,\gamma)$ cross sections for $^{129-132}$Sn for neutron energies of 5 and 10 MeV appears
due to the overall shift of the cross sections toward higher energies, similar as observed
in Fig. \ref{Sn-isotopes} for $(n,\alpha)$ reaction.

\subsection{Maxwellian averaged cross sections along Fe and Sn isotope chains}\label{Sec4}

Since $(n,\alpha)$ reactions also contribute to the nucleosynthesis,
it is interesting to explore the isotopic dependence of the cross
sections in stellar environment. Only small modifications of the neutron capture reaction cross sections could have important implications 
on the path of nuclear processes contributing to the synthesis of
chemical elements \cite{Dan2019}. We investigate the
Maxwellian averaged cross sections (MACS), that take into account 
Maxwell-Boltzmann distribution for neutrons with respect to the corresponding temperature \cite{Pritychenko2010}.
In the present study, two temperatures characteristic for 
stellar environment are considered, $kT=\text{30 keV}$ (e.g., associated to
the core He burning in massive stars), and $kT=\text{90 keV}$ (e.g., in  supernova envelope where the r-process could occur) \cite{Pignatari2010, Janka2007,Arnould2007}.

So far, available information on the MACS values for $(n,\alpha)$ reactions
with Fe and Sn isotopes is rather limited. At 30 keV, for ${}^{46-78}$Fe isotopes the only  calculations are available for ${}^{55}$Fe \cite{ROSFOND2008}, ${}^{57}$Fe \cite{JEFF3.1} and ${}^{59}$Fe \cite{ROSFOND2008}. For ${}^{100-132}$Sn isotopes at 30 keV, there are also only a few available MACS values. For ${}^{112,113,115,117}$Sn, there are a few calculations \cite{ENDFBVII-0, ENDFB7.1, JEFF3.1, JENDL3-3, ROSFOND2008, JENDL4.0, CENDL3.1}, but for ${}^{121}$Sn \cite{ROSFOND2008} and ${}^{114,119,123,125,126}$Sn \cite{JENDL4.0, JEFF3.1} there are results only from one calculation. At 90 keV, there are MACS values for ${}^{54, 55, 56, 57, 59}$Fe in Refs. \cite{JENDL3-3, JENDL4.0, CENDL3.1, ROSFOND2008, ENDFBVII-0, ENDFB7.1, JEFF3.1}. For ${}^{112,113,114,115,117}$Sn a few results exist in Refs. \cite{ENDFBVII-0,ENDFB7.1,JEFF3.1,JENDL3-3,JENDL4.0,ROSFOND,CENDL3.1}. 
For ${}^{119,121,123,125}$Sn there exists only one result for each isotope \cite{JENDL4.0, ROSFOND2008, JEFF3.1}.
Thus we perform a systematic calculation of the MACS values over
Fe and Sn isotope chains. For comparison, in addition to the $(n,\alpha)$
reaction, also other exit channels from the neutron induced reactions 
are studied.

\begin{figure}
\centering
\includegraphics[width=\linewidth]{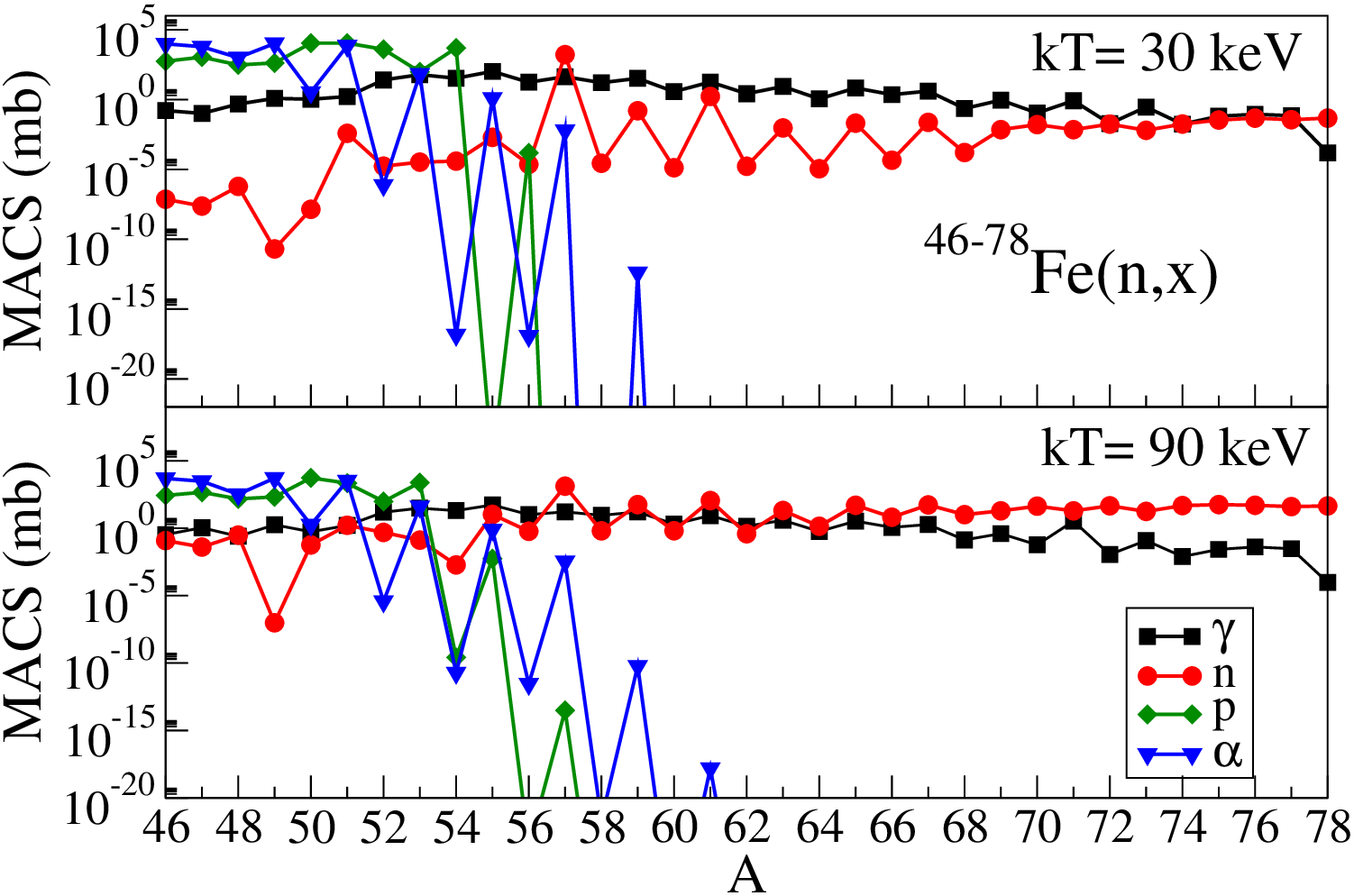}
\caption{The Maxwellian averaged cross sections (MACS) for neutron induced reactions with ${}^{46-78}$Fe at $kT= \text{30 keV}$ and 90 keV, with $\alpha$, $\gamma$, proton and neutron emission.}\label{Fe-MACS-30}
\end{figure}

Figure \ref{Fe-MACS-30} shows the Maxwellian averaged cross sections 
for ${}^{46-78}$Fe isotopes with temperature corresponding to
$kT= \text{30 keV}$ (top panel) and $kT= \text{90 keV}$ (lower panel), 
for the neutron induced reactions with $\gamma$, $n$, $p$, and $\alpha$ emission from compound nucleus.
For $kT= \text{30 keV}$, for lighter Fe isotopes one can observe that $\alpha$ emission has large contributions comparable as those with $\gamma$, $n$ and $p$ emission. Especially for ${}^{46,47,48,49,51}$Fe, the cross section for $\alpha$ exit channel is the most dominant,
and for ${}^{50,53,55}$Fe it also has relevant contribution. As expected from the previous analysis of the cross sections, the MACS values also rapidly decrease for the neutron-rich isotopes. One can also observe the
sensitivity of the results on temperature, resulting in the reduction
of the MACS values and modifications in their evolution along the isotope chains.

\begin{figure}
\centering
\includegraphics[width=\linewidth]{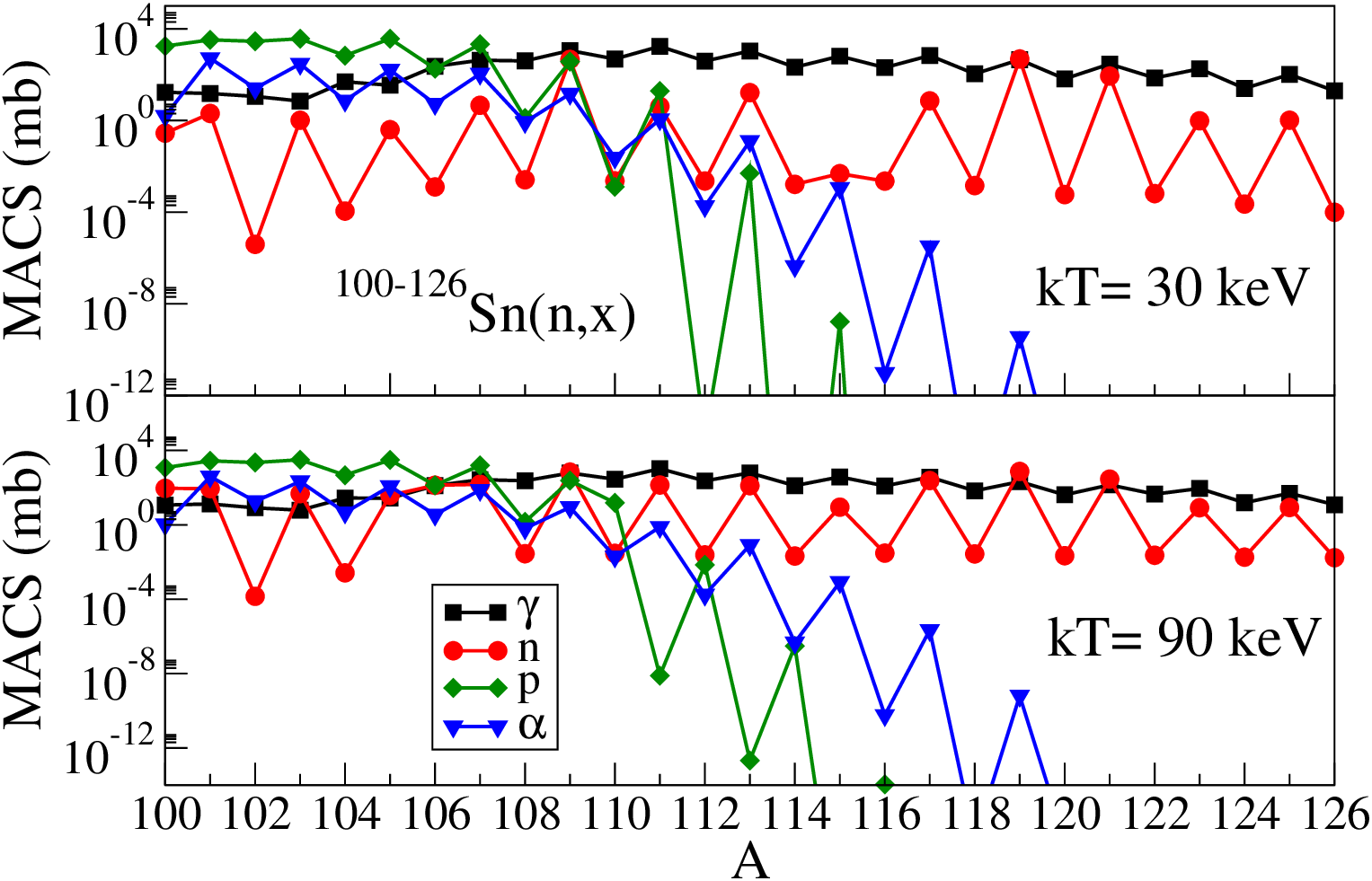}
\caption{The same as Fig.\ref{Fe-MACS-30}, but for ${}^{100-126}$Sn.}\label{Sn-MACS-30}
\end{figure}

Similar analysis of the MACS values is also performed for ${}^{100-126}$Sn at $kT= \text{30 keV}$ and $kT= \text{90 keV}$. In Fig. \ref{Sn-MACS-30} the corresponding MACS results are shown separately for $\gamma$, $n$, $p$, and $\alpha$ emission in the exit channels. While the MACS values for $\alpha$ emission are decreasing and becoming negligible for target nuclei with larger number of neutrons, for isotopes ${}^{100-107}$Sn the $\alpha$ emission also contributes, though not as dominant channel. Further more systematic studies,
along the nuclide map, are required to explore in more details the
contributions of nuclei with considerable $(n,\alpha)$ reaction cross 
sections to the nucleosynthesis.

 \section{Conclusion}\label{Sec5}
In this work we have investigated the evolution of $(n,\alpha)$ reaction cross sections for target nuclei within Fe and Sn isotope chains. Model calculations have been performed in the theory framework based on Hauser-Feshbach statistical model and exciton model through their implementation in the nuclear reaction code TALYS \cite{Talys}, using Skyrme
functional to describe nuclear masses and back-shifted Fermi gas model level densities. 
Global optical model potential has been used, with
additionally adjusted rescaling of the mass dependent geometrical 
parameters, 
obtained by $\chi^2$ minimization using experimental data 
for $^{54}$Fe and $^{118}$Sn to benchmark the model for studies
of $(n,\alpha)$ reactions along the respective isotope chains.
In this way, the present model calculations resulted in 
considerable improvement compared to previous studies.
It remains open question how reliable are adjusted
optical model potentials for isotopes away from ${}^{54}$Fe and ${}^{118}$Sn used to improve the description of $(n,\alpha)$ reaction cross 
sections. The available data on ${}^{56,58}$Fe and ${}^{112}$Sn are well reproduced, however, more experimental studies on $(n,\alpha)$ reaction cross sections are required to assess the quality of the optical model potentials when extrapolating to neutron-rich isotopes.

Since the experimental data on $(n,\alpha)$ reaction cross sections
are rather limited for Fe and Sn isotopes, and often restricted to narrow energy range, model calculations in this work provide insight
how the cross sections vary with increasing neutron number, by considering
complete relevant neutron energy range.
It is shown that $(n,\alpha)$ reaction cross section have larger
values for lighter isotopes for both isotope chains, and their
maximal values are obtained for $^{53}$Fe and $^{103}$Sn,
respectively. By further increasing the neutron number in 
target nucleus, the $(n,\alpha)$ reaction cross section 
becomes smaller due to decreasing of the reaction $Q$-value and
opening other more dominant exit channels,
as shown in the analysis that in addition to $\alpha$
emission included also neutron, proton, and $\gamma$ emission. 
The neutron energy has an effective role for the
$(n,\alpha)$ cross section, i.e., neutrons with higher
energies can induce reactions in heavier isotopes. 
The analysis of Maxwellian averaged cross sections showed
dominant or considerable contributions of $(n,\alpha)$ reactions  
over other exit channels for lower mass Fe isotopes. 
Throughout the isotope chains, while going to neutron-rich isotopes, 
cross section values strongly decrease. 
Thus, the $(n,\alpha)$ reactions are not effective for 
very neutron rich nuclei, and their role in the 
nucleosynthesis that goes through the neutron-rich side 
of the nuclide map, such as the r-process, is negligible.
While the $(n,\alpha)$ reactions dominate for low-mass Fe
isotopes, in nuclei with neutron excess $\gamma$ and neutron
emission are dominating outcomes from neutron induced reactions,
with varying contributions dependent on temperature involved. 
Further studies are required to assess the role of
$(n,\alpha)$ reactions in nucleosynthesis, in cases
of nuclei with significant respective cross sections.  

\section{Acknowledgements}
This work is supported by the QuantiXLie Centre of Excellence, a project co-financed by the Croatian Government and European Union through the European Regional Development Fund, the Competitiveness and Cohesion Operational Programme (KK.01.1.1.01.0004). 
S.K. acknowledges support from the Scientific and Technological Research Council of Turkey (TUBITAK) through the International Doctoral Research Fellowship Programme 2214A, 2020/1, Grant No. 1059B142000254.

\bibliographystyle{elsarticle-harv}
\bibliography{nalpha.bbl}
\end{document}